\documentclass[12pt,preprint]{aastex}

\begin{document}

\title{Relativistic equation of state at subnuclear densities
       in the Thomas-Fermi approximation}

\author{Z. W. Zhang and H. Shen}
\affil{School of Physics, Nankai University, Tianjin 300071, China}
\email{shennankai@gmail.com}

\begin{abstract}
We study the non-uniform nuclear matter using the self-consistent
Thomas--Fermi approximation with a relativistic mean-field model.
The non-uniform matter is assumed to be
composed of a lattice of heavy nuclei surrounded by dripped nucleons.
At each temperature $T$, proton fraction $Y_p$, and baryon mass density $\rho_B$,
we determine the thermodynamically favored state by minimizing the free energy
with respect to the radius of the Wigner--Seitz cell,
while the nucleon distribution in the cell can be determined
self-consistently in the Thomas--Fermi approximation.
A detailed comparison is made between the present results and
previous calculations in the Thomas--Fermi approximation with a parameterized
nucleon distribution that has been adopted in the widely used Shen EOS.
\end{abstract}


\section{Introduction}
\label{sec:1}

The equation of state (EOS) of hot and dense matter is an essential ingredient
in understanding many astrophysical phenomena, e.g., supernova explosions and
neutron star formations~\citep{burr06,jank07,sumi05,sumi09,shen11}.
The EOS for the core-collapse supernova simulations must cover wide ranges
of temperature, proton fraction, and baryon density (see Table 1 of~\citet{shen11}).
Therefore, it is very difficult to build a complete EOS covering the wide
range of thermodynamic conditions.
Many efforts have been made to investigate the EOS of
nuclear matter for the use of supernova simulations and neutron star
calculations~\citep{latt91,latt07,shen98a,shen98b,shen11,scha96,webe05}.
There are two commonly used EOSs in supernova simulations,
namely the Lattimer--Swesty EOS~\citep{latt91}, which employed
a compressible liquid-drop model with a Skyrme force,
and the Shen EOS~\citep{shen98b,shen11}, which used a relativistic
mean-field (RMF) model and Thomas--Fermi approximation with a
parameterized nucleon distribution.
Recently, \citet{shen10} constructed the EOS based on a relativistic
Hartree calculation for the Wigner--Seitz cell which includes nuclear shell effects.
These EOSs employ the so-called single nucleus approximation (SNA),
in which only a single representative nucleus is included instead of a
distribution of different nuclei.
It would be desirable to consider the mixture of nuclei based on
nuclear statistical equilibrium~\citep{hemp10,blin11,furu11,furu13}.
The mixture of nuclei is important for electron captures on nuclei inside
supernova core.
However, it has been demonstrated that SNA is a reasonable approximation
for thermodynamical quantities~\citep{burr84}.
In SNA, the thermodynamically favored nucleus is described by a compressible
liquid-drop model in the Lattimer--Swesty EOS or by a Thomas--Fermi
approximation with parameterized nucleon distribution in the Shen EOS.
In this paper, we intend to study the matter at subnuclear densities,
in which the heavy nucleus is described by a self-consistent Thomas--Fermi
approximation.

The self-consistent Thomas--Fermi approximation has been widely used in
atomic and nuclear physics.
Many properties of nuclei can be described by the Thomas--Fermi approximation
with good agreement to experimental data~\citep{TF07}.
Recently, the self-consistent Thomas--Fermi approximation has been used to
study nuclear pasta phases at subnuclear densities
at zero temperature~\citep{Avancini09} and finite temperature~\citep{Avancini10},
where the pasta phases include droplets (bubbles), rods (tubes), and slabs
for three, two, and one dimensions, respectively.
In our previous work~\citep{shen98a,shen98b,shen11}, a parameterized
nucleon distribution was assumed in the Thomas--Fermi approximation
and only droplet phase was taken into account.
It is, however, not clear how good/bad the assumed nucleon distribution
functions are in Shen EOS. Also, whether other pasta phases, like bubble
phase, can make a meaningful difference in the transition to uniform
nuclear matter. The main purpose of the present work is to
study the non-uniform matter at subnuclear densities using the self-consistent
Thomas--Fermi approximation. By comparing the nucleon distributions and
thermodynamic quantities, we can examine the difference between the
self-consistent Thomas--Fermi (STF) approximation and
the parameterized Thomas--Fermi (PTF) approximation.
In the present work, we consider both droplet and bubble configurations
in order to investigate the effect of including the bubble phase,
while other pasta phases are neglected for simplicity.

For the effective nuclear interaction, we use the relativistic mean-field (RMF)
theory, in which nucleons interact via the exchange of isoscalar scalar and
vector mesons ($\sigma$ and $\omega$) and an isovector vector meson ($\rho$).
In this work, we employ the RMF theory including nonlinear $\sigma$ and $\omega$
terms with the parameter set TM1~\citep{suga94}. It is known that
the RMF theory with the parameter set TM1 can well reproduce
ground state properties of finite nuclei including unstable ones~\citep{suga94},
and predicts a maximum neutron-star mass of $2.18\ M_\odot$~\citep{shen11}.
In Shen EOS, the RMF results of TM1
were taken as input in the PTF calculation.
Therefore, a detailed comparison can be made between the STF
and PTF approximations based on the same RMF theory.

This paper is organized as follows. In Section~\ref{sec:2}, we briefly
explain the RMF theory and the STF approximation
for the non-uniform matter at subnuclear densities.
In Section~\ref{sec:3}, we discuss the calculated results of
the STF approximation in comparison with those obtained in the PTF approximation.
Section~\ref{sec:4} is devoted to the conclusions.

\section{Formalism}
\label{sec:2}

We first give a brief description of the RMF theory~\citep{sero86,suga94}.
We employ the RMF theory to calculate the properties of uniform matter.
For non-uniform matter where nuclei exist to decrease the free energy,
we use the STF approximation in which the RMF Lagrangian is used to
derive the equations of motion for the fields~\citep{Avancini09}.

In the RMF theory, nucleons interact via the exchange of mesons.
The exchanged mesons are isoscalar scalar and vector mesons ($\sigma$ and $\omega$)
and isovector vector meson ($\rho$). We adopt the RMF theory with
nonlinear $\sigma$ and $\omega$ terms~\citep{suga94}.
For a system consisting of protons, neutrons, and electrons,
the Lagrangian density reads,
\begin{eqnarray}
\label{eq:LRMF}
{\cal L}_{\rm{RMF}} & = & \sum_{i=p,n}\bar{\psi}_i\left[i\gamma_{\mu}\partial^{\mu} -M
-g_{\sigma}\sigma-g_{\omega}\gamma_{\mu}\omega^{\mu}
-g_{\rho}\gamma_{\mu}\tau_a\rho^{a\mu}
-e \gamma_{\mu}\frac{1+\tau_3}{2} A^{\mu}
\right]\psi_i  \nonumber\\
& & +\bar{\psi}_{e}\left[i\gamma_{\mu}\partial^{\mu} -m_{e}
+e \gamma_{\mu} A^{\mu} \right]\psi_{e}  \nonumber\\
 && +\frac{1}{2}\partial_{\mu}\sigma\partial^{\mu}\sigma
-\frac{1}{2}m^2_{\sigma}\sigma^2-\frac{1}{3}g_{2}\sigma^{3}
-\frac{1}{4}g_{3}\sigma^{4} \nonumber\\
 && -\frac{1}{4}W_{\mu\nu}W^{\mu\nu}
+\frac{1}{2}m^2_{\omega}\omega_{\mu}\omega^{\mu}
+\frac{1}{4}c_{3}\left(\omega_{\mu}\omega^{\mu}\right)^2   \nonumber\\
 && -\frac{1}{4}R^a_{\mu\nu}R^{a\mu\nu}
+\frac{1}{2}m^2_{\rho}\rho^a_{\mu}\rho^{a\mu}
-\frac{1}{4}F_{\mu\nu}F^{\mu\nu},
\end{eqnarray}
where $W^{\mu\nu}$, $R^{a\mu\nu}$, and $F^{\mu\nu}$ are the antisymmetric
field tensors for $\omega^{\mu}$, $\rho^{a\mu}$, and $A^{\mu}$, respectively.
We use the parameter set TM1~\citep{suga94} as give in Table~\ref{tab:1}.
It is known that the RMF theory with the parameter set TM1 can reproduce
good saturation properties of nuclear matter and satisfactory description
of finite nuclei~\citep{suga94,hira96}.

Starting with the Lagrangian~(\ref{eq:LRMF}), we derive a set of
Euler--Lagrange equations.
In the RMF approximation, the meson fields are considered as classical fields
and they are replaced by their expectation values.
For a static system, the non-vanishing expectation values
are   $\sigma =\left\langle \sigma    \right\rangle$,
      $\omega =\left\langle \omega^{0}\right\rangle$,
      $\rho   =\left\langle \rho^{30} \right\rangle$,
  and $ A     =\left\langle  A^{0}\right\rangle$.
The equations of motion for these mean fields have the following form:
\begin{eqnarray}
\label{eq:ms0}
& &-\nabla^2\sigma + m_\sigma^2\sigma +g_{2}\sigma^{2}+g_{3}\sigma^{3} =
 -g_{\sigma} n_s,
\\
\label{eq:mw0}
& & -\nabla^2\omega + m_\omega^2\omega +c_3\omega^3  =
  g_{\omega} n_v,
\\
\label{eq:mr0}
& & -\nabla^2\rho + m_\rho^2\rho  =
 g_{\rho} n_3,
\\
\label{eq:A0}
& & -\nabla^2 A    =
  e n_c,
\end{eqnarray}
where $n_s$, $n_v$, $n_3$, and $n_c$
are the scalar, vector, third component of isovector,
and charge densities, respectively.
The stationary Dirac equation for nucleons is given by
\begin{eqnarray}
\label{eq:driacn}
 & & \left(-\mathbf{\alpha}\cdot \mathbf{\nabla} + \beta M^{*}
+g_{\omega}\omega +g_{\rho}\tau_3\rho
+e \frac{1+\tau_3}{2} A \right)\psi^{i}=
\varepsilon^{i} \psi^{i} ,
\end{eqnarray}
where $M^{*}=M+g_{\sigma}\sigma$ is the effective nucleon mass.
$i$ denotes the index of eigenstates, while $\varepsilon^{i}$
is the single-particle energy.

For non-uniform matter at subnuclear densities, heavy nuclei exist
in order to decrease the free energy.
We assume that each spherical nucleus is
located in the center of a charge-neutral cell consisting of a
vapor of nucleons and electrons.
In the present study, we focus on estimating the difference between
the STF and PTF approximations,
so we ignore the contribution of alpha-particles for simplicity.
The alpha-particle fraction has been shown in Figure 6 of~\citet{shen11},
and moreover the contributions from alpha-particles and other light nuclei
have been extensively discussed in~\citet{sumi08} and~\citet{hemp10}.
We assume that nuclei are arranged in a body-centered-cubic (BCC) lattice to
minimize the Coulomb lattice energy~\citep{oyam93}. The Wigner--Seitz cell is introduced
to simplify the energy of a unit cell, which is a sphere with the same volume
as the unit cell in the BCC lattice. The lattice constant $a$ and
the radius of the Wigner--Seitz cell $R_C$ are related to
the cell volume by $ V_{\rm{cell}}=a^3=4 \pi R_C^3 / 3=N_B / n_B $,
where $N_B$ and $n_{B}$ are the baryon number per cell and the average baryon
number density, respectively.

In the STF approximation,
the nucleon distribution function at position $r$
inside the Wigner--Seitz cell is obtained by
\begin{equation}
\label{eq:nirmf}
 n_{i}(r)=\frac{1}{\pi^2}
       \int_0^{\infty} dk\,k^2\,\left[f_{i}^{k}(r)-f_{\bar{i}}^{k}(r)\right],
\end{equation}
where $f_{i}^{k}$ and $f_{\bar{i}}^{k}$ ($i=p$, $n$) are the occupation probabilities
of the particle and antiparticle for momentum $k$.
At zero temperature, $f_{i}^{k}=1$ under the Fermi surface
and $f_{i}^{k}=0$ above the Fermi surface.
At finite temperature, the occupation probability is obtained by the Fermi--Dirac
distribution,
\begin{eqnarray}
\label{eq:fp}
 f_{i}^{k}=\frac{1}{1+\exp\left[\left(\sqrt{k^2+{M^{*}}^2}-\nu_{i}\right)
        /T\right]},
\\
\label{eq:fa}
 f_{\bar{i}}^{k}=\frac{1}{1+\exp\left[\left(\sqrt{k^2+{M^{*}}^2}
 +\nu_{i}\right)/T\right]}.
\end{eqnarray}
The chemical potential $\mu_i$ is related to the effective chemical
potential $\nu_i$ as
\begin{eqnarray}
\label{eq:mup}
 \mu_{p}       &=& \nu_p      +g_{\omega}\omega +g_{\rho}\rho + e A, \\
\label{eq:mun}
 \mu_{n}       &=& \nu_n      +g_{\omega}\omega -g_{\rho}\rho.
\end{eqnarray}
We note that the chemical potential is spatially constant throughout the Wigner--Seitz
cell, while other quantities such as occupation probabilities and mean-field values
depend on the position $r$.  As for the electrons, we disregard the electron
screening effect caused by the non-uniform charged particle distributions,
and assume the electron density is uniform.
It was found in~\citet{Maru05} that the electron screening effect is very small
at subnuclear densities. For given average baryon density $n_B$ and proton
fraction $Y_p$, the electrons do not play any role in the free energy minimization,
therefore, we ignore the electron contribution as done in~\citet{shen11}.

The free energy per cell contributed from baryons is given by
\begin{equation}
\label{eq:Fcell}
F_{\rm{cell}}=E_{\rm{cell}}- T S_{\rm{cell}},
\end{equation}
where $E_{\rm{cell}}$ and $S_{\rm{cell}}$ denote the energy and entropy
per cell, respectively. The energy per cell can be written as
\begin{eqnarray}
E_{\rm{cell}} &=&\int_{\rm{cell}} \epsilon (r) d^3r
+ \Delta E_C,
\label{eq:Ecell}
\end{eqnarray}
with $\Delta E_C$ being the correction term for the BCC
lattice~\citep{oyam93,shen11}.
This correction is negligible when the nuclear size
is much smaller than the cell size.
The entropy per cell is given by
\begin{eqnarray}
S_{\rm{cell}} &=&\int_{\rm{cell}} s (r) d^3r.
\label{eq:Scell}
\end{eqnarray}
Here $\epsilon (r)$ and $s(r)$ are the local energy density and entropy density
at the radius $r$, which can be calculated by using the RMF theory
and the STF approximation.
The energy density in the STF approximation is given by
\begin{eqnarray}
\label{eq:ETF}
\epsilon &=& \displaystyle{\sum_{i=p,n} \frac{1}{\pi^2}
   \int_0^{\infty} dk\,k^2\,
   \sqrt{k^2+{M^*}^2}  \left(f_{i}^{k}+f_{\bar{i}}^{k}\right) }
   \nonumber\\  & &
 +\frac{1}{2}(\nabla \sigma )^{2}
 +\frac{1}{2}m_{\sigma}^2\sigma^2+\frac{1}{3}g_{2}\sigma^{3}+\frac{1}{4}g_{3}\sigma^{4}
 \nonumber\\  & &
 -\frac{1}{2}(\nabla \omega )^{2}-\frac{1}{2}m_{\omega}^2\omega^2-\frac{1}{4}c_{3}\omega^{4}
 +g_{\omega}\omega \left(n_p+n_n\right)
 \nonumber\\  & &
\label{eq:STF}
 -\frac{1}{2}(\nabla \rho )^{2}-\frac{1}{2}m_{\rho}^2\rho^2
 +g_{\rho}\rho \left(n_p-n_n\right)
  \nonumber\\  & &
 -\frac{1}{2}(\nabla A)^{2}+e A \left(n_p-n_e\right)  ,
\end{eqnarray}
the entropy density is given by
\begin{eqnarray}
s & =  \displaystyle{\sum_{i=p,n} \frac{1}{\pi^2} \int_0^{\infty} dk\,k^2 }
   & \left[ -f_{i}^{k}\ln f_{i}^{k}
            -\left(1-f_{i}^{k}\right)\ln \left(1-f_{i}^{k}\right) \right. \nonumber\\
 & & \left. -f_{\bar{i}}^{k}\ln f_{\bar{i}}^{k}
            -\left(1-f_{\bar{i}}^{k}\right)\ln \left(1-f_{\bar{i}}^{k}\right) \right] .
\end{eqnarray}
Here, we have omitted the electron kinetic energy and electron entropy,
which do not play any role in the free energy minimization.
It is known that nuclear pasta phases could present at subnuclear densities
before the transition to uniform matter.
In this study, we consider both droplet and bubble phases.
The free energy for bubble phase can also be calculated from Equation~(\ref{eq:Fcell}),
but with different correction term $\Delta E_C$~\citep{oyam93}.

It is interesting and important to compare the difference between the
STF used in this study and the PTF adopted in~\citet{shen11}.
In the PTF approximation, the energy per cell is given in the form
\begin{eqnarray}
E^{\rm{PTF}}_{\rm{cell}} &=& E_{\rm{cell}}^b+E_{\rm{cell}}^g+E_{\rm{cell}}^C,
\label{eq:EPTF}
\end{eqnarray}
where $E_{\rm{cell}}^b$ is the bulk energy per cell given
by Equation (20) of~\citet{shen11}.
The surface (gradient) energy term $E_{\rm{cell}}^g$ due to the inhomogeneity
of the nucleon distribution
is assumed to have the form
\begin{equation}
E_{\rm{cell}}^g=\int_{\rm{cell}} F_0 \mid \nabla \left[ \, n_n\left(r\right)+
    n_p\left(r\right) \, \right] \mid^2 d^3r,
\label{eq:ES}
\end{equation}
with the parameter $F_0=70 \, \rm{MeV\,fm^5}$ determined by the gross
properties of nuclear masses and charge radii~\citep{oyam93,shen11}.
The Coulomb energy term $E_{\rm{cell}}^C$ can be calculated as
\begin{equation}
\label{eq:EC}
E_{\rm{cell}}^C=\frac{1}{2}\int_{\rm{cell}} e A\left(r\right)
                \left[n_p\left(r\right)-n_e\right] d^3r
               + \Delta E_C,
\end{equation}
where $A\left(r\right)$ is the electrostatic potential and
$\Delta E_C$ is the same as that of Equation~(\ref{eq:Ecell}).
By comparing Equation~(\ref{eq:EPTF}) with Equation~(\ref{eq:Ecell}),
we recognize that the Coulomb energy in STF and PTF
can be calculated in the same way as given by Equation~(\ref{eq:EC}),
but the surface (gradient) energy is treated differently.
In the STF approximation the gradient energy is included in
Equation~(\ref{eq:ETF}) self-consistently, while it is calculated
by Equation~(\ref{eq:ES}) with an additional parameter $F_0$ in the PTF approximation.
This may cause some differences in energy between these two methods.
Another difference between STF and PTF is that
the nucleon distribution function $n_i(r)$ ($i=p$ or $n$)
is determined self-consistently in the STF approximation by solving
Equations~(\ref{eq:ms0})-(\ref{eq:A0}) inside the Wigner--Seitz cell.
In the PTF method, the nucleon distribution function is assumed
to have the form~\citep{oyam93,shen11}
\begin{equation}
\label{eq:nitf}
n_i\left(r\right)=\left\{
\begin{array}{ll}
\left(n_i^{\rm{in}}-n_i^{\rm{out}}\right) \left[1-\left(\frac{r}{R_i}\right)^{t_i}
\right]^3 +n_i^{\rm{out}},  & 0 \leq r \leq R_i, \\
n_i^{\rm{out}},  & R_i \leq r \leq R_C. \\
\end{array} \right.
\end{equation}
It is important to examine the effect of these differences on thermodynamic
quantities at subnuclear densities, so that we can estimate how good/bad
the PTF approximation is in~\citet{shen11}.

We minimize the free energy per baryon, $F=F_{\rm{cell}}/N_B$,
at given temperature $T$, proton fraction $Y_p$,
and baryon mass density $\rho_B$. Note that the baryon mass density is defined
as $\rho_B=m_{u} n_B$ with $m_{u}$ being the atomic mass unit~\citep{shen11}.
The thermodynamically favored state is the one with the lowest $F$
among all configurations considered. In the PTF approximation,
the minimization procedure was realized with respect to
several independent parameters as described in~\citet{shen11}.
In the STF approximation, we minimize $F$ with respect to
the Wigner--Seitz cell radius, $R_C$, and finally determine the most stable
configuration among droplet, bubble, and homogeneous phases by comparing
their free energies.
To compute $F_{\rm{cell}}$ at a fixed $R_C$, we numerically solve the
coupled Equations~(\ref{eq:ms0})-(\ref{eq:A0}) together with
the nucleon distribution and occupation probability given
by Equations~(\ref{eq:nirmf})-(\ref{eq:fa})
in coordinate space.
Starting from an initial guess for the mean-field values $\sigma (r)$,
$\omega (r)$, $\rho (r)$, and $A(r)$, we determine
the chemical potential $\mu_i$ ($i=p$ or $n$) by the condition
$\int_{0}^{R_C} n_i (r) 4\pi r^2 dr=N_i$,
where the proton and neutron numbers per cell are respectively given by
$N_p=Y_p N_B$ and $N_n=(1-Y_p) N_B$.
Once the chemical potentials are known, the occupation probabilities
and density distributions can be obtained. Then, using these densities
we solve Equations~(\ref{eq:ms0})-(\ref{eq:A0})
to get new mean-field values. This procedure is iterated until
self-consistency is achieved.
For the numerical integrations in coordinate space,
we use a composite Simpson's rule with 1201 grid points.
In general, this number of grid points is sufficient
to achieve good convergence in both solving coupled equations and
performing numerical integrations.

\section{Results and discussion}
\label{sec:3}

In this section, we show and discuss the results of the non-uniform matter
at subnuclear densities obtained using the STF approximation.
In comparison with the PTF method used in~\citet{shen11},
the nucleon distribution and the surface effect
are self-consistently calculated within the STF approximation.
In addition, we take into account the bubble phase that may
be present before the transition to uniform matter.

\begin{figure}[htb]
\begin{center}
\includegraphics[bb=15 15 550 770, width=8.6 cm,clip]{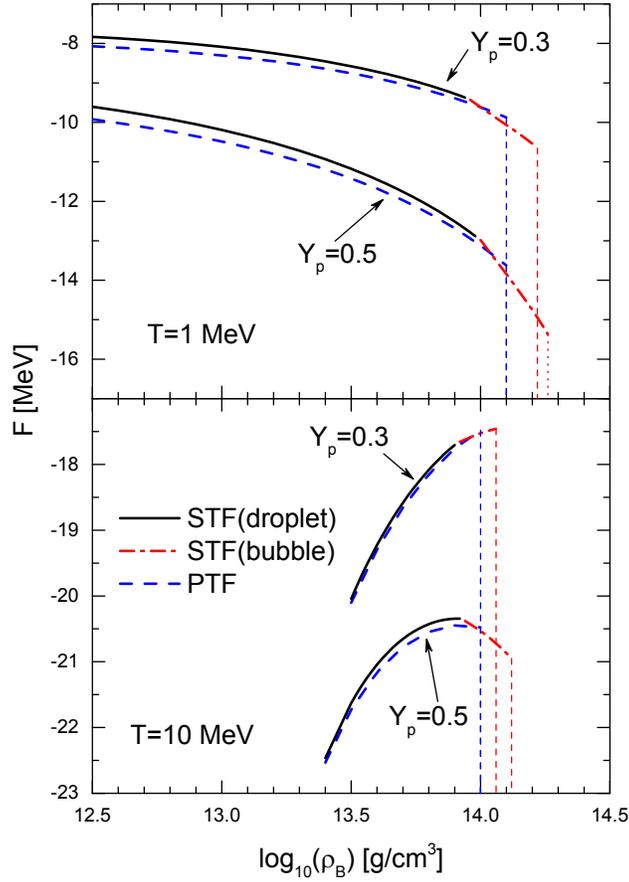}
\caption{Free energy per baryon $F$ versus
$\rho_B$ for $Y_p=0.3$ and $0.5$ at $T=1$ MeV (top panel)
and $T=10$ MeV (bottom panel).
The results of STF with droplet configuration (black solid lines)
and bubble configuration (red dash-dotted lines)
are compared with those of PTF (blue dashed lines).
The vertical dashed lines indicate the position where
the transition from non-uniform matter to uniform matter occurs.
(A color version of this figure is available in the online journal.)}
\label{fig:1F}
\end{center}
\end{figure}

At given temperature $T$, proton fraction $Y_p$,
and baryon mass density $\rho_B$, we minimize the free energy per
baryon, $F=F_{\rm{cell}}/N_B$, with respect to the independent
variables in the model. In the STF approximation, there is only
one independent variable, namely the cell radius $R_C$.
However, there are about seven independent variables
($n_n^{\rm{in}}$, $R_n$, $t_n$, $n_p^{\rm{in}}$, $R_p$, $t_p$, and $R_C$)
in the PTF approximation~\citep{shen11}.
It is interesting and important to make a detailed comparison
between STF and PTF.
In Figure~\ref{fig:1F}, we show the resulting free energy per
baryon $F$ versus the baryon mass density $\rho_B$
for $Y_p=0.3$ and $0.5$ at $T=1$ MeV and $10$ MeV.
Note that we focus here on the non-uniform matter phase in which
heavy nuclei are formed in the medium-density and low-temperature region,
while the behavior of $F$ and other thermodynamic quantities
over the wide range of EOS has been discussed in our earlier
work~\citep{shen98b,shen11}.
We present in Figure~\ref{fig:1F} the results of STF (PTF)
with a droplet configuration by black solid (blue dashed) lines.
The bubble phase is also taken into account in the STF calculation
as shown by red dash-dotted lines.
It is shown that the onset density of the bubble phase
is above $10^{13.9}\,\rm{g\,cm^{-3}}$.
The inclusion of the bubble phase causes a visible decrease
in the free energy at $\rho_B > 10^{13.9}\,\rm{g\,cm^{-3}}$.
On the other hand, the appearance of the bubble phase can delay
the transition to uniform matter as indicated by
the vertical dashed lines.
We find that there is a small difference in $F$ between STF and PTF,
especially in the case of $T=1$ MeV (top panel).
The free energy per baryon $F$ obtained in PTF
is systematically lower than that of STF
for the same droplet configuration.
This may be due to the different treatment of surface effect and
nucleon distribution between these two methods.
In order to estimate how much difference can be caused by the different
treatment of surface effect, we should compare corresponding terms
in Equation~(\ref{eq:EPTF}).
However, it is difficult in the STF approximation to separate
the gradient energy from the bulk energy, because both of them
are involved in the first term of Equation~(\ref{eq:Ecell}).
On the other hand, the Coulomb energy can be easily separated from
Equation~(\ref{eq:Ecell}) as defined by Equation~(\ref{eq:EC}),
so that it is possible to compare the difference in the Coulomb energy
between STF and PTF.
In~\citet{oyam93} and~\citet{oyam03}, the authors have pointed out that
the gradient energy in equilibrium should be as large as the Coulomb
energy in general cases, which means that
$E_{\rm{cell}}^g \simeq E_{\rm{cell}}^C$ could hold in both STF and PTF.
This relation corresponds to the well-known equilibrium condition
in the liquid-drop model that the surface energy is twice as much
as the Coulomb energy.
In the results of PTF, we do obtain
$E_{\rm{cell}}^g = E_{\rm{cell}}^C$ (see Table~\ref{tab:2}).
Therefore, we can use the relation $E_{\rm{cell}}^g = E_{\rm{cell}}^C$
to estimate the gradient energy in the STF approximation,
and define the bulk energy as
$E_{\rm{cell}}^b=E_{\rm{cell}}-E_{\rm{cell}}^g-E_{\rm{cell}}^C$.
In Table~\ref{tab:2}, we compare various quantities between STF
and PTF for the cases of $Y_p=0.3$ and $T=1$ MeV.
The definitions of these quantities are as follows:
$F=F_{\rm{cell}}/N_B$ is the free energy per baryon,
$E=E_{\rm{cell}}/N_B$ is the energy per baryon,
$S=S_{\rm{cell}}/N_B$ is the entropy per baryon,
$E_b=E_{\rm{cell}}^b/N_B$ is the bulk energy per baryon,
$E_g=E_{\rm{cell}}^g/N_B$ is the gradient energy per baryon,
$E_C=E_{\rm{cell}}^C/N_B$ is the Coulomb energy per baryon,
and $R_C$ is the radius of the Wigner--Seitz cell.
Note that $F_0=70 \, \rm{MeV\,fm^5}$ has been used in the PTF
method~\citep{shen11}, so we first compare the results of STF
with those of PTF ($F_0=70$).
It is shown that there is no much difference in $S$ and $E_b$,
but $F$, $E$, $E_g$, and $E_C$ of PTF ($F_0=70$)
are all slightly lower than those of STF.
Furthermore, the difference in $F$
(which is $0.217$ MeV at $\rho_B = 10^{13.0}\,\rm{g\,cm^{-3}}$)
is about twice as much as that in $E_C$ ($\sim 0.1$ MeV).
This implies that the difference in $F$ is mostly caused by the
sum $E_g + E_C = 2 E_g $, namely the surface effect.
It seems that $E_g$ with $F_0=70 \, \rm{MeV\,fm^5}$ in the PTF
approximation is relatively small compared to the self-consistent
calculation of STF. To analyze the influence of the parameter $F_0$,
we recalculated the results of PTF with $F_0=90 \, \rm{MeV\,fm^5}$,
that are also listed in Table~\ref{tab:2}.
By comparing the results of PTF ($F_0=70$) and PTF ($F_0=90$),
we find that $E_g$ and $E_C$ of PTF ($F_0=90$) are significantly
enhanced and closer to the values of STF.
As a results, the differences in $F$ between STF and PTF ($F_0=90$)
are much smaller than those between STF and PTF ($F_0=70$).
Since $E_g$ and $E_C$ of PTF ($F_0=90$) are very close to the values of STF,
the small differences in $F$ between STF and PTF ($F_0=90$)
should be caused by the different treatment of nucleon distributions
between these two methods.
In the last column of Table~\ref{tab:2}, we compare the cell radius
$R_C$ obtained by different methods. It is shown that $R_C$ of
PTF ($F_0=70$) is obviously smaller than that of STF and PTF ($F_0=90$).
This is because a smaller surface energy favors a smaller nuclear size
and cell radius based on the liquid-drop model~\citep{Maru05}.
Therefore, the increase of the surface energy in PTF ($F_0=90$)
leads to a larger $R_C$ compared to that in PTF ($F_0=70$).
In the bottom panel of Figure~\ref{fig:1F}, we show the results
for the case of $T=10$ MeV.
The density range of the non-uniform matter phase at
high temperature becomes very narrow as shown in Figure 2 of~\citet{shen11},
so we just compare the results in this density range.
It is seen that the differences between STF and PTF at $T=10$ MeV (bottom panel)
are generally smaller than those at $T=1$ MeV (top panel).
This is because at higher temperature the entropy becomes more
dominant and the treatment of surface effect plays a less important
role in determining the free energy.

\begin{figure}[htb]
\begin{center}
\includegraphics[bb=15 15 550 780, width=8.6 cm,clip]{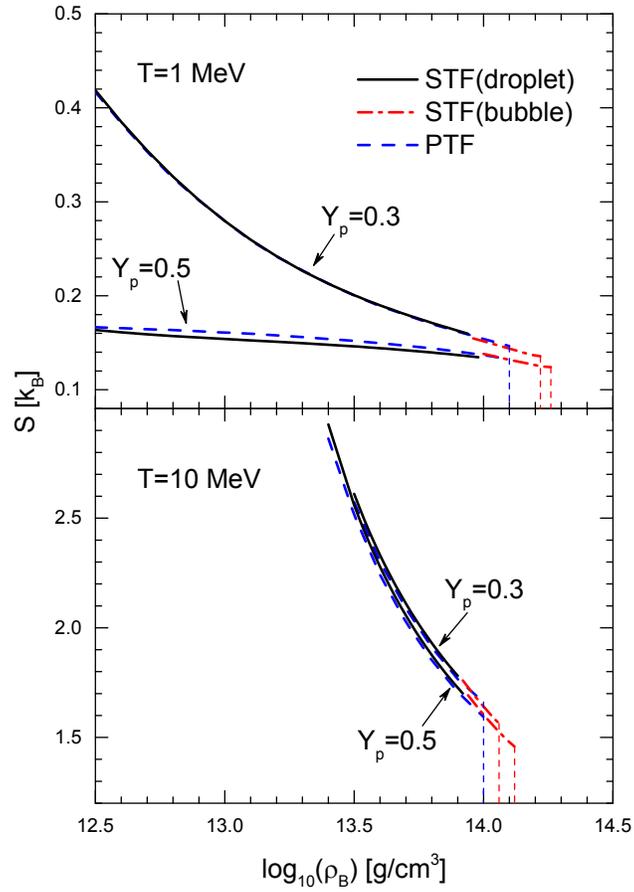}
\caption{Same as Figure~\ref{fig:1F}, but for entropy per baryon $S$.
(A color version of this figure is available in the online journal.)}
\label{fig:2S}
\end{center}
\end{figure}

We plot in Figure~\ref{fig:2S} the entropy per baryon $S$ versus
$\rho_B$ for $Y_p=0.3$ and $0.5$ at $T=1$ MeV and $10$ MeV.
In the case of $T=1$ MeV (top panel), there is almost no difference
between STF and PTF for $Y_p=0.3$ (also see Table~\ref{tab:2}),
while there is a small difference in the case of
$Y_p=0.5$. At $T=10$ MeV (bottom panel), the results of STF and PTF
are almost identical for both $Y_p=0.3$ and $Y_p=0.5$.
We note that the behavior of the entropy has a strong $Y_p$ dependence,
which is due to the formation of heavy nuclei as discussed in~\citet{shen98b,shen11}.

\begin{figure}[htb]
\begin{center}
\includegraphics[bb=15 15 560 760, width=8.6 cm,clip]{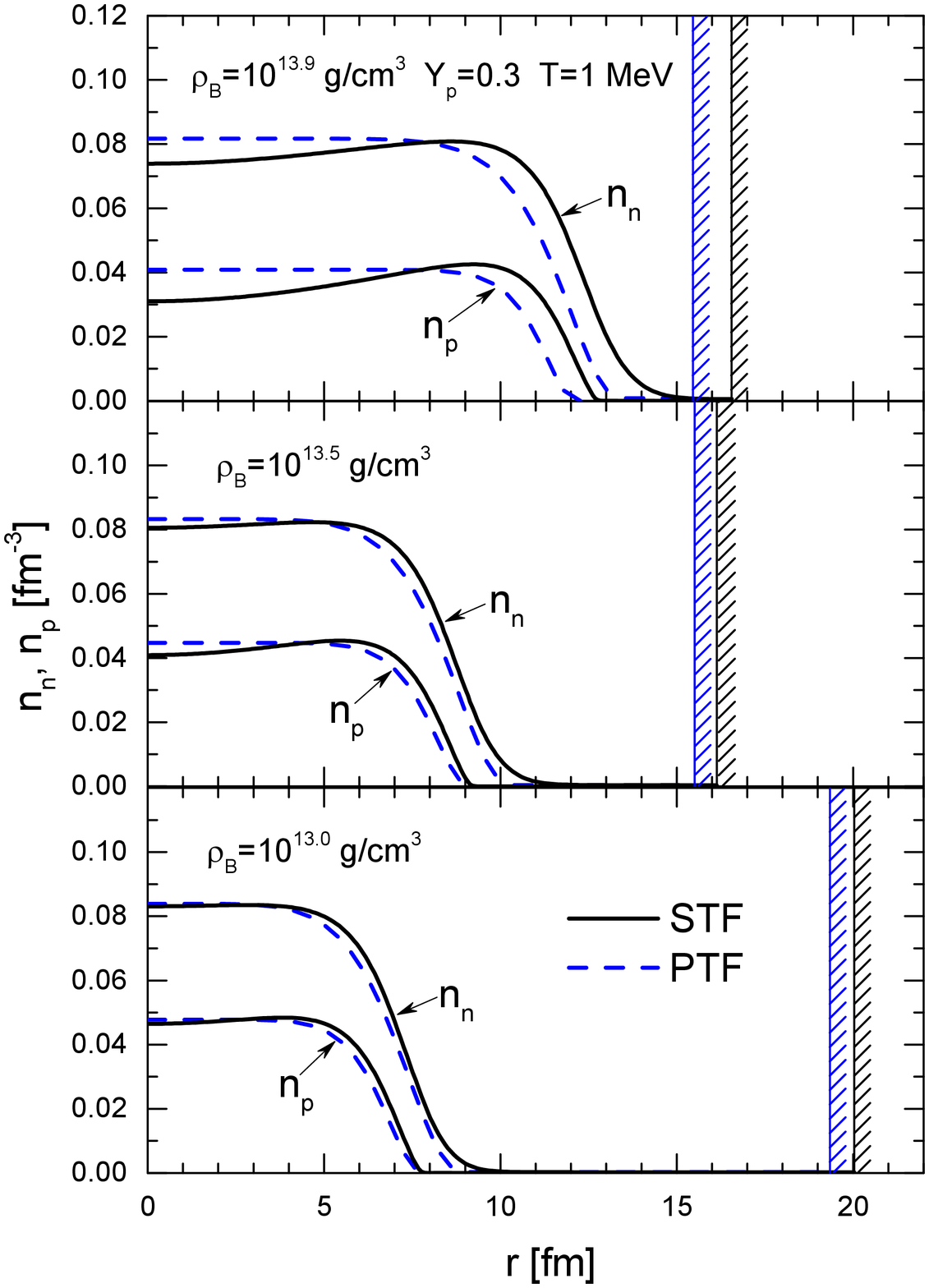}
\caption{Density distributions of protons and neutrons inside the
Wigner--Seitz cell for the cases of $Y_p=0.3$ and $T=1$ MeV
at $\rho_B = 10^{13.0},\, 10^{13.5}$, and $10^{13.9}\, \rm{g\,cm^{-3}}$ (bottom to top).
The cell radius is indicated by the hatch.
The results of STF (black solid lines)
are compared with those of PTF (blue dashed lines).
(A color version of this figure is available in the online journal.)}
\label{fig:3DT1}
\end{center}
\end{figure}

\begin{figure}[htb]
\begin{center}
\includegraphics[bb=15 15 550 730, width=8.6 cm,clip]{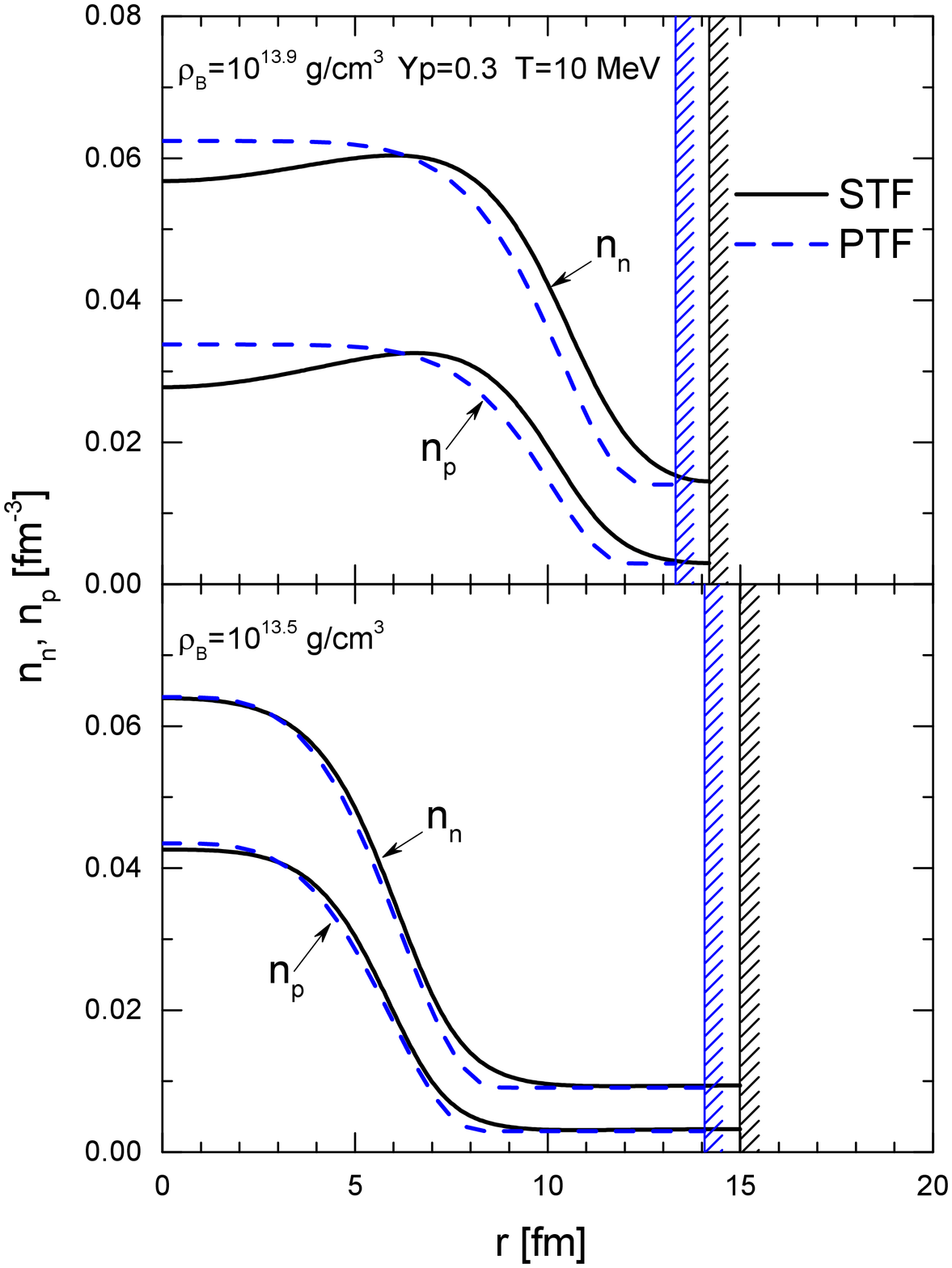}
\caption{Density distributions of protons and neutrons inside the
Wigner--Seitz cell for the cases of $Y_p=0.3$ and $T=10$ MeV
at $\rho_B = 10^{13.5}$ and $10^{13.9}\, \rm{g\,cm^{-3}}$ (bottom to top).
The cell radius is indicated by the hatch.
The results of STF (black solid lines)
are compared with those of PTF (blue dashed lines).
(A color version of this figure is available in the online journal.)}
\label{fig:4DT10}
\end{center}
\end{figure}

In Figures~\ref{fig:3DT1} and~\ref{fig:4DT10}, we show the
density distributions of protons and neutrons inside the
Wigner--Seitz cell for the cases of $Y_p=0.3$ at $T=1$ MeV
and $T=10$ MeV, respectively.
The horizontal axis denotes the radial distance from the center
of the cell, while the cell radius is indicated by the hatch.
The results obtained in STF (black solid lines)
are compared with those of PTF (blue dashed lines).
At lower densities, there is no obvious difference in the density
profiles between STF and PTF. However, as density increases,
the difference becomes noticeable as shown in the top panels.
It is seen that the densities at the center of the cell
are significantly lower than those at the surface region obtained
in the STF approximation. This is because the Coulomb interaction
is explicitly included in the equation of motion for the protons,
and as a result, more protons are pushed off to the surface.
The same behavior has been observed in~\citet{Maru05}
where the authors compare results obtained with different
treatment of the Coulomb interaction.
In the STF approximation, the nucleon distributions are
obtained self-consistently with the cell radius $R_C$
determined by the free energy minimization.
However, the nucleon distributions in the PTF approximation
are forced to have the form of Equation~(\ref{eq:nitf})
with all parameters including $R_C$ determined
in the minimization procedure.
Comparing the results of $T=10$ MeV (Figure~\ref{fig:4DT10})
with those of $T=1$ MeV (Figure~\ref{fig:3DT1}),
the differences between STF and PTF
are very similar. It is shown that more free nucleons exist
outside the nuclei in the case of $T=10$ MeV.
This is because at higher temperature the entropy becomes
more dominant, and as a result, the free energy could be
decreased by more nucleons dripping out of the nuclei.
In both Figures~\ref{fig:3DT1} and~\ref{fig:4DT10},
the cell radius $R_C$ obtained in STF is obviously
larger than that of PTF. This is related to the treatment of
surface effect as discussed above. It is known based on the
liquid-drop model that a smaller surface energy favors a smaller
nuclear size and cell radius~\citep{Maru05}.
In the PTF method, $F_0=70 \, \rm{MeV\,fm^5}$
has been used in the calculation of surface (gradient) energy,
which seems not large enough in comparison with the results
of STF (see Table~\ref{tab:2}).
Therefore, the smaller surface energy of PTF leads to
a smaller $R_C$ compared to that of STF.

\begin{figure}[htb]
\begin{center}
\includegraphics[bb=10 15 550 800, width=8.6 cm,clip]{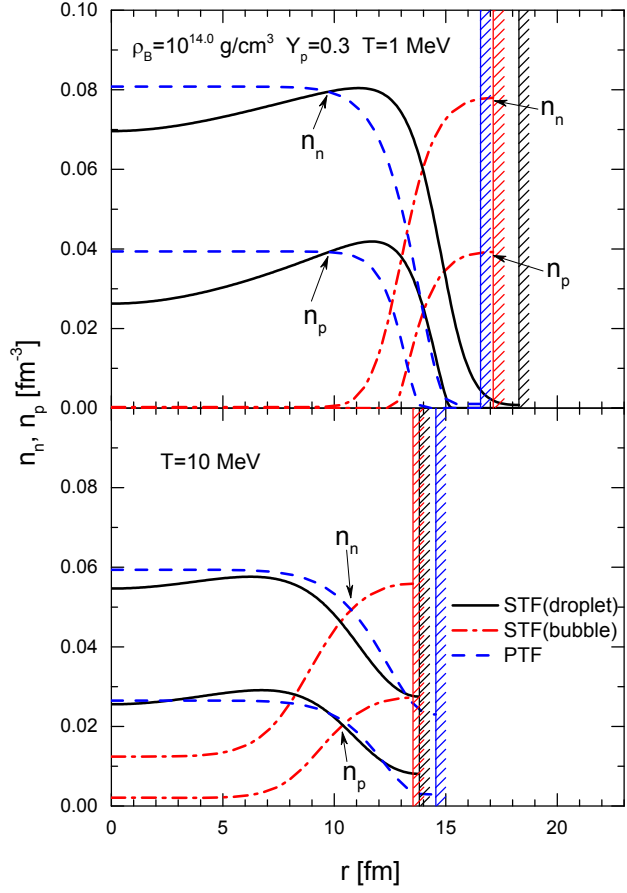}
\caption{Density distributions of protons and neutrons inside the
Wigner--Seitz cell for the cases of $Y_p=0.3$ and
$\rho_B = 10^{14.0}\, \rm{g\,cm^{-3}}$ at
$T=1$ MeV (top panel) and $T=10$ MeV (bottom panel).
The cell radius is indicated by the hatch.
The red dash-dotted lines display the results of STF with bubble configuration.
The black solid lines illustrate the results of STF with droplet configuration,
while the blue dashed lines show those of PTF for comparison.
(A color version of this figure is available in the online journal.)}
\label{fig:5D}
\end{center}
\end{figure}

We consider both droplet and bubble phases in this study.
It is found that the bubble could have a lower free energy
than the droplet near the transition density to uniform matter.
In Figure~\ref{fig:5D}, we show the density distributions
of protons and neutrons obtained with droplet and bubble configurations
using the STF approximation for the cases of $Y_p=0.3$ and
$\rho_B = 10^{14.0}\,\rm{g\,cm^{-3}}$ at $T=1$ MeV (top panel)
and $T=10$ MeV (bottom panel).
We minimize the free energy per baryon with respect to the
cell radius for both droplet and bubble configurations,
and then determine the most stable droplet and bubble.
By comparing their free energies, we determine
which is the most favorable configuration among droplet, bubble,
and homogeneous phases.
The onset of the bubble phase can be seen in Figure~\ref{fig:1F}.
Generally, the bubble has the lowest free energy at
$\rho_B \sim 10^{14}\,\rm{g\,cm^{-3}}$ ($n_B \sim 0.06\,\rm{fm^{-3}}$).
We present in Table~\ref{tab:3} the resulting properties of stable droplet
and bubble in the STF approximation,
while the results of PTF and those of uniform matter
are also listed for comparison.
It is shown that the difference in $F$
between the droplet and bubble phases is less than $1\%$,
but there are significant differences in $\mu_p$ and $\mu_n$.
On the other hand, the inclusion of the bubble phase can increase
the transition density to uniform matter. For instance,
at $\rho_B = 10^{14.2}\,\rm{g\,cm^{-3}}$ with $Y_p=0.3$ and $T=1$ MeV,
the bubble phase has the lowest free energy among droplet, bubble,
and homogeneous phases, but it favors the homogeneous phase
if the bubble configuration is not taken into account.

\begin{figure}[htb]
\begin{center}
\includegraphics[bb=10 20 550 770, width=8.6 cm,clip]{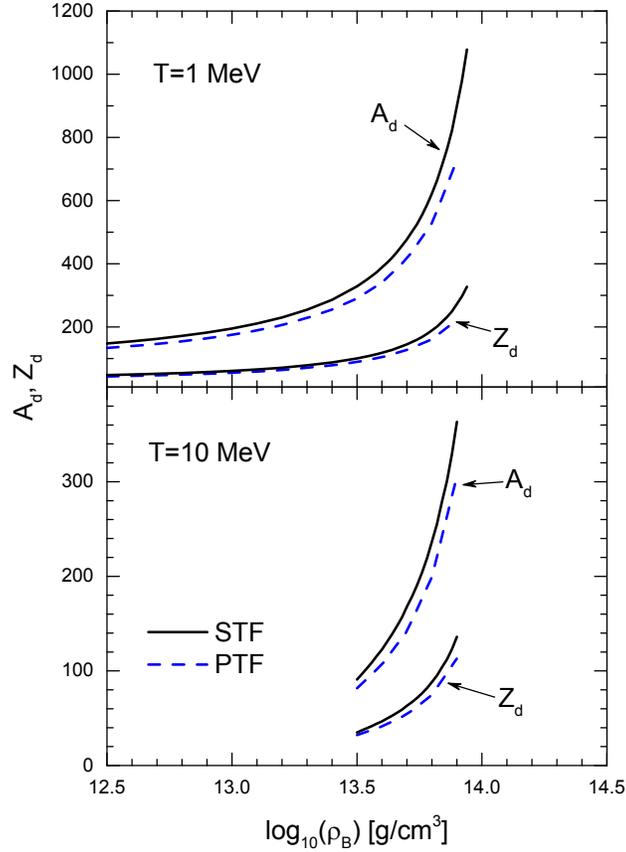}
\caption{Nuclear mass number $A_d$ and charge number $Z_d$
inside the droplet as a function of $\rho_B$
for $Y_p=0.3$ at $T=1$ MeV (top panel) and $T=10$ MeV (bottom panel).
Note that $A_d$ and $Z_d$ are defined as excesses with respect to
the background nucleon gas using the subtraction procedure.
The results of STF (black solid lines)
are compared with those of PTF (blue dashed lines).
(A color version of this figure is available in the online journal.)}
\label{fig:6AZ}
\end{center}
\end{figure}

\begin{figure}[htb]
\begin{center}
\includegraphics[bb=15 15 550 780, width=8.6 cm,clip]{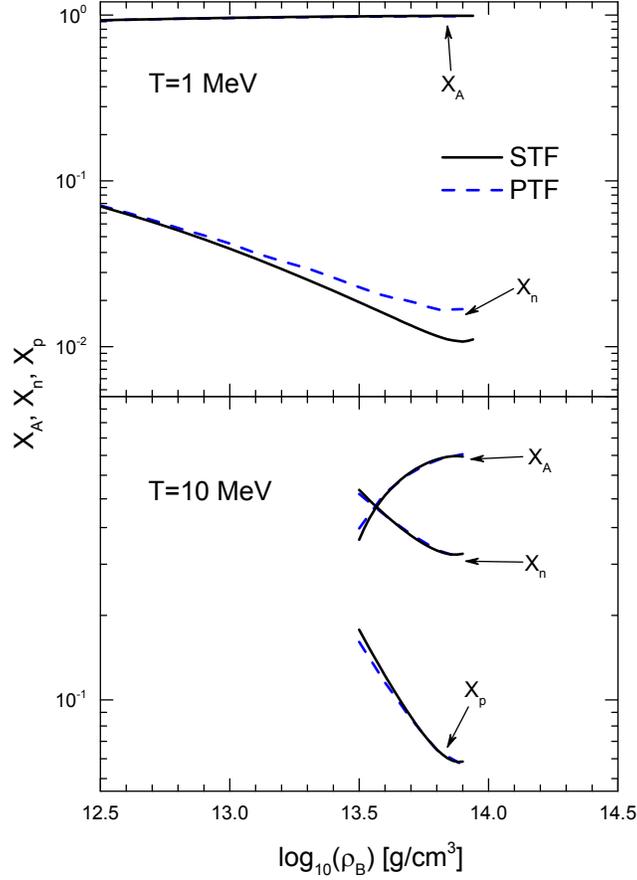}
\caption{Fractions of nuclei ($X_A$), neutron gas ($X_n$),
and proton gas ($X_p$) as a function of $\rho_B$
for $Y_p=0.3$ at $T=1$ MeV (top panel) and $T=10$ MeV (bottom panel).
Nuclei are defined as excesses with respect to
the background nucleon gas using the subtraction procedure.
The results of STF (black solid lines)
are compared with those of PTF (blue dashed lines).
(A color version of this figure is available in the online journal.)}
\label{fig:7Yi}
\end{center}
\end{figure}

We examine the droplet properties in non-uniform matter
and investigate their density dependence.
In Figure~\ref{fig:6AZ}, we show the nuclear mass number $A_d$
and charge number $Z_d$ inside the droplet as a function of
$\rho_B$ for the cases of $Y_p=0.3$ at $T=1$ MeV (top panel)
and $T=10$ MeV (bottom panel).
Note that these quantities are different from those shown in Figure 5
of~\citet{shen11}. Here the background nucleon gas is subtracted
in order to isolate the nucleus from the surrounding nucleon gas,
namely $A_d=N_B-V_{\rm{cell}} n_B(R_C)$ and $Z_d=N_p-V_{\rm{cell}} n_p(R_C)$.
This subtraction procedure has been widely used in Thomas--Fermi
calculations~\citep{De01,Gril12}. For comparison, we calculate
$A_d$ and $Z_d$ using the PTF approximation and show with blue dashed lines.
It is seen that $A_d$ and $Z_d$ increase rapidly with increasing density.
At the same $\rho_B$ and $Y_p$, the values of $A_d$ and $Z_d$ at $T=10$
MeV are significantly less than those at $T=1$ MeV,
which is due to more nucleons can drip out of the nuclei at higher temperature.
It is found that there is a small difference between STF and PTF
for both $T=1$ MeV and $T=10$ MeV. This should be related to
the difference in nucleon distributions as shown in Figures~\ref{fig:3DT1}
and~\ref{fig:4DT10}. Generally, the droplet properties obtained in
STF are very similar to those of PTF.
In Figure~\ref{fig:7Yi}, we show the fractions of nuclei ($X_A$), neutron gas ($X_n$),
and proton gas ($X_p$) as a function of $\rho_B$
for the same case as Figure~\ref{fig:6AZ}.
These fractions are defined by
$X_A=A_d / N_B$, $X_n=V_{\rm{cell}} n_n(R_C)/N_B$, and $X_p=V_{\rm{cell}} n_p(R_C)/N_B$.
In the case of $T=1$ MeV and $Y_p=0.3$ (top panel),
there is almost no proton gas ($X_p \simeq 0$), while the neutron gas fraction $X_n$
is very small and decreases with increasing density.
This implies that nucleons inside the droplet are dominant at low temperature.
For the case of $T=10$ MeV and $Y_p=0.3$ (bottom panel),
more nucleons can drip out of the nuclei as shown in Figures~\ref{fig:4DT10},
and as a result $X_n$ is of the same order of $X_A$,
while $X_p$ is about one order lower than $X_n$.
Comparing the results between STF and PTF, it is hard to see any significant
difference at $T=10$ MeV (bottom panel),
while there is a small difference in $X_n$ at $T=1$ MeV (top panel).

\begin{figure}[htb]
\begin{center}
\includegraphics[bb=55 430 510 760, width=15.0 cm,clip]{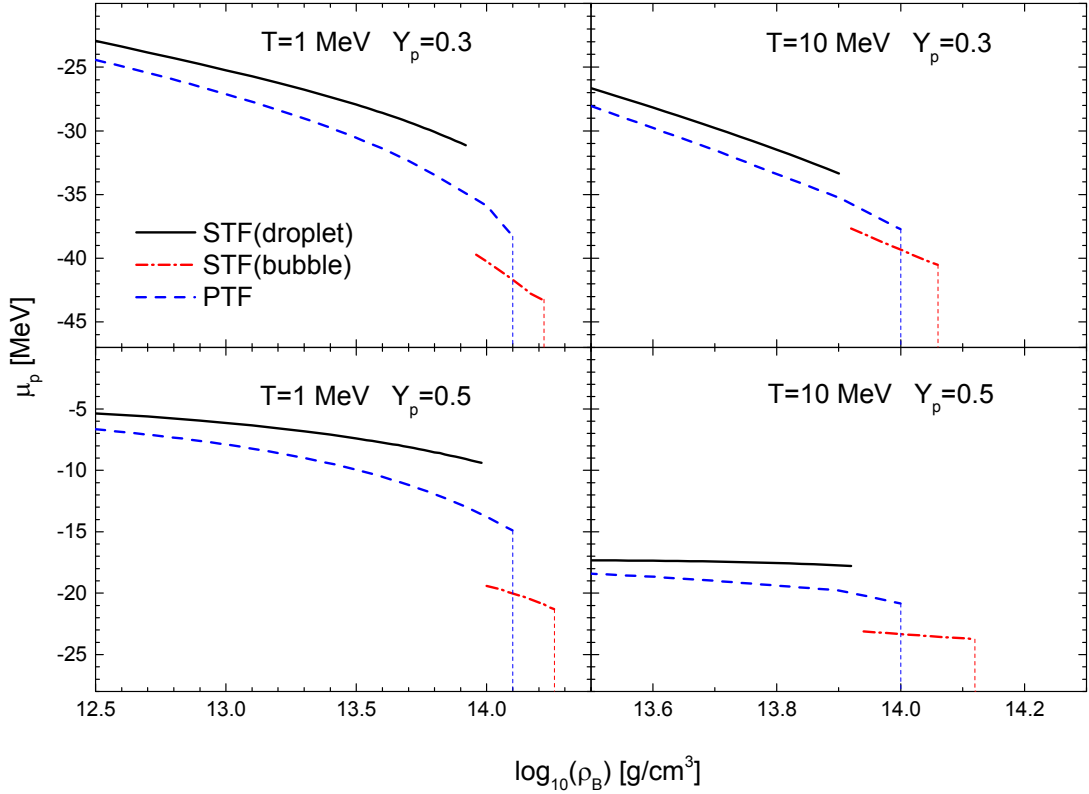}
\caption{Proton chemical potential $\mu_p$ as a function of
$\rho_B$ for $Y_p=0.3$ (top panels) and $Y_p=0.5$ (bottom panels)
at $T=1$ MeV (left panels) and $T=10$ MeV (right panels).
The results of STF with droplet configuration (black solid lines)
and bubble configuration (red dash-dotted lines)
are compared with those of PTF (blue dashed lines).
The onset of the bubble phase at $\rho_B > 10^{13.9}\,\rm{g\,cm^{-3}}$
causes sudden jumps in $\mu_p$ of STF.
The vertical dashed lines indicate the position where
the transition from non-uniform matter to uniform matter occurs.
(A color version of this figure is available in the online journal.)}
\label{fig:8Mup}
\end{center}
\end{figure}

\begin{figure}[htb]
\begin{center}
\includegraphics[bb=55 430 510 760, width=15.0 cm,clip]{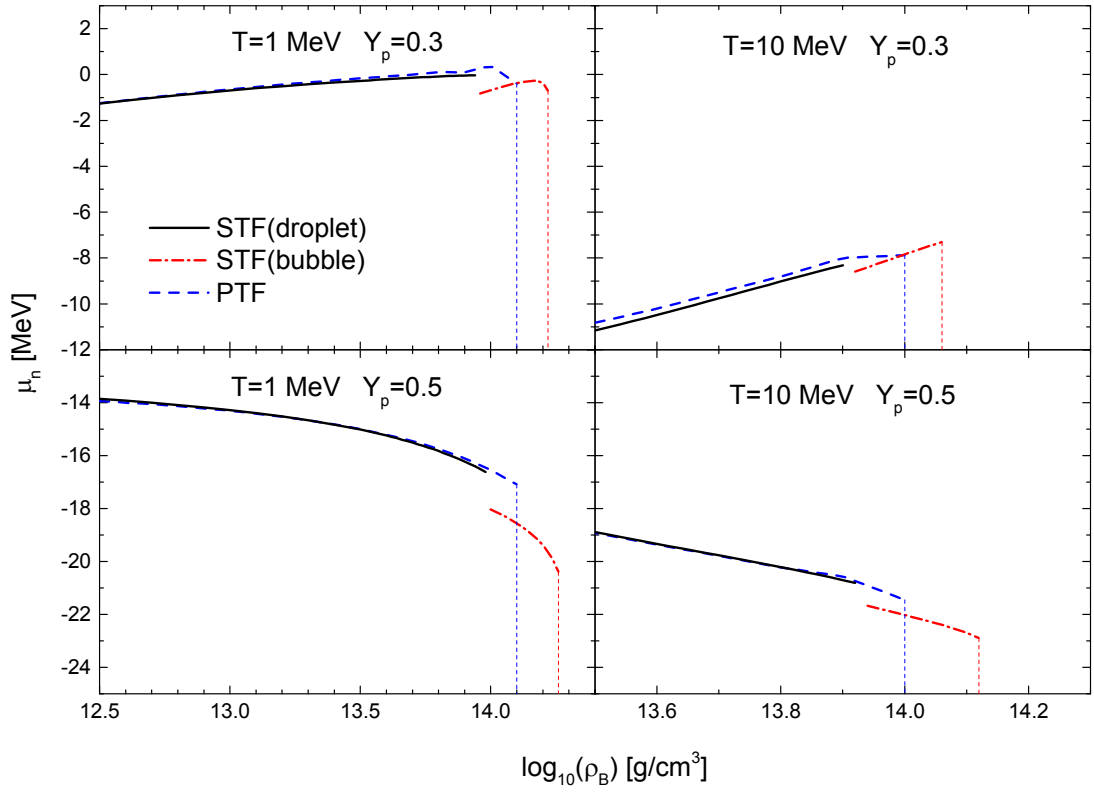}
\caption{Same as Figure~\ref{fig:8Mup}, but for neutron chemical potential $\mu_n$.
(A color version of this figure is available in the online journal.)}
\label{fig:9Mun}
\end{center}
\end{figure}

In Figures~\ref{fig:8Mup} and~\ref{fig:9Mun}, we show the chemical potentials
of protons and neutrons, $\mu_p$ and $\mu_n$, as a function of $\rho_B$
with $Y_p=0.3$ and $0.5$ at $T=1$ MeV and $10$ MeV.
The results of PTF are taken from EOS2 of~\citet{shen11},
which were calculated through the thermodynamic relations given in Equations~(A16)
and (A17) of~\citet{shen11}.
In the STF approximation, the chemical potentials given in
Equations~(\ref{eq:mup}) and~(\ref{eq:mun}) are obtained
self-consistently as described in Section~\ref{sec:2},
which are spatially constant throughout the Wigner--Seitz cell.
It is shown that the appearance of the bubble phase
at $\rho_B > 10^{13.9}\,\rm{g\,cm^{-3}}$ causes sudden jumps
in $\mu_p$ and $\mu_n$ within the STF approximation.
This is mainly because the Coulomb potential in the bubble is
very different from that in the droplet.
As for comparison between STF and PTF,
it is found that there are visible differences between STF and PTF
in $\mu_p$ as shown in Figure~\ref{fig:8Mup}, while the chemical
potentials of neutrons are almost identical between STF and PTF
with the same droplet configuration as shown in Figure~\ref{fig:9Mun}.
The difference in $\mu_p$ may be related to the difference in Coulomb
interaction between STF and PTF.
As discussed above, the Coulomb and surface energies in PTF
with $F_0=70 \, \rm{MeV\,fm^5}$ is relatively small
compared to those of STF, which means that the Coulomb potential
in PTF should be smaller than that in STF.
According to Equation~(\ref{eq:mup}), a larger Coulomb potential
corresponds to a higher $\mu_p$. Therefore, we obtain higher $\mu_p$
in STF due to its larger Coulomb potential.
On the other hand, $\mu_n$ is not directly related to the Coulomb potential,
so the difference in $\mu_n$ between STF and PTF is very small as shown
in Figure~\ref{fig:9Mun}.

\section{Conclusion}
\label{sec:4}

In this paper, we have studied the non-uniform matter at subnuclear densities
using the STF approximation. For the effective nuclear interaction, we have
adopted the RMF theory including nonlinear $\sigma$ and $\omega$ terms
with the parameter set TM1 which can reproduce good saturation properties
of nuclear matter and satisfactory description of finite nuclei.
We have made a detailed comparison
between the STF approximation used in this study and the PTF approximation
adopted in~\citet{shen11}.
In addition, we have included the bubble phase that could be present
before the transition to uniform matter. It has been found that
the inclusion of the bubble phase can significantly affect the chemical
potentials of protons and neutrons, while its effects on free energy
and entropy are relatively small. Furthermore, the appearance of the
bubble phase can delay the transition to uniform matter.

We have examined the difference between STF and PTF.
In the STF method, the nucleon distribution and the surface effect are
treated self-consistently. We have minimized the free energy with respect to
the cell radius at given temperature $T$, proton
fraction $Y_p$, and baryon mass density $\rho_B$.
The thermodynamically favored state is the one with the lowest free energy
among all configurations considered.
The results obtained in the STF approximation have been compared with
those of PTF. It have been found that there is
no obvious difference in nucleon distributions at lower densities,
while the difference becomes noticeable near the transition density
to uniform matter. For thermodynamical quantities, such as the free
energy and entropy per baryon, the results of both methods generally
agree well with each other. However, there are some small differences
between STF and PTF which need to be analyzed.
The free energy per baryon obtained in PTF is slightly lower than
that of STF for the same droplet configuration.
This is mainly caused by the inconsistent treatment of the surface
effect in PTF, namely the surface and Coulomb
energies with the parameter $F_0=70 \, \rm{MeV\,fm^5}$ is
relatively small compared to those obtained self-consistently in STF.
In addition, the smaller surface energy in PTF leads to a smaller
cell radius in comparison to that of STF. On the other hand,
the proton chemical potential obtained in STF is slightly higher
than that of PTF, which is also related to the difference
in the Coulomb and surface energies between STF and PTF.
Therefore, we can draw the conclusion that most of the differences
between STF and PTF should be due to the different treatment
of surface effect, namely the parameter $F_0$ used in PTF is not
large enough in comparison with the results obtained in the
STF approximation.

Considering the wide range of thermodynamic conditions in the whole
EOS~\citep{shen11}, the differences between STF and PTF
are thought to be negligible and cannot affect the general behavior
of the EOS. Therefore, we conclude that the PTF approximation is a
reasonable description for non-uniform matter, and can produce
very similar EOS with that obtained in the STF approximation
which is considered to be self-consistent in the treatment
of surface effect and nucleon distribution.

\acknowledgments

This research is supported in part by the National Natural
Science Foundation of China (Grants No. 11075082 and No. 11375089).


\clearpage
\begin{deluxetable}{cccccccccc}
\tabletypesize{\scriptsize}
\tablecaption{Parameter set TM1 for the RMF Lagrangian used in this work.
              The masses are given in $\rm{MeV}$.
\label{tab:1}}
\tablewidth{0cm}
\tablehead{
 \colhead{$M$} & \colhead{$m_\sigma$} & \colhead{$m_\omega$} & \colhead{$m_\rho$} &
 \colhead{$g_\sigma$} & \colhead{$g_\omega$} & \colhead{$g_\rho$} &
 \colhead{$g_2$ ($\rm{fm}^{-1}$)} & \colhead{$g_3$} & \colhead{$c_3$}
 }
\startdata
 938.0 & 511.19777 & 783.0 & 770.0 & 10.02892 & 12.61394 & 4.63219 & -7.23247 & 0.61833 & 71.30747 \\
\enddata
\end{deluxetable}


\begin{deluxetable}{clccccccc}
\tabletypesize{\scriptsize}
\tablecaption{
Comparison between different methods for the
cases of $Y_p=0.3$ and $T=1$ MeV
at $\rho_B = 10^{13.0},\, 10^{13.5}$, and $10^{13.9}\, \rm{g\,cm^{-3}}$.
The various quantities are defined as follows:
$F=F_{\rm{cell}}/N_B$ is the free energy per baryon,
$E=E_{\rm{cell}}/N_B$ is the energy per baryon,
$S=S_{\rm{cell}}/N_B$ is the entropy per baryon,
$E_b=E_{\rm{cell}}^b/N_B$ is the bulk energy per baryon,
$E_g=E_{\rm{cell}}^g/N_B$ is the gradient energy per baryon,
$E_C=E_{\rm{cell}}^C/N_B$ is the Coulomb energy per baryon,
and $R_C$ is the radius of the Wigner--Seitz cell.
\label{tab:2}}
\tablewidth{0cm}
\tablehead{
 \colhead{$\log_{10}(\rho_B)$} & \colhead{method} & \colhead{$F$} & \colhead{$E$}& \colhead{$S$}     & \colhead{$E_b$} & \colhead{$E_g$} & \colhead{$E_C$} & \colhead{$R_c$} \\
 \colhead{($\rm{g\,cm^{-3}}$)} & \colhead{}   & \colhead{(MeV)} & \colhead{(MeV)}& \colhead{($k_B$)} & \colhead{(MeV)} & \colhead{(MeV)} & \colhead{(MeV)} & \colhead{(fm)}
}
\startdata
 13.0 & STF            & -8.087 & -7.807 & 0.280 & -10.135 & 1.164 & 1.164 & 20.0\\
      & PTF ($F_0=70$) & -8.304 & -8.025 & 0.278 & -10.161 & 1.068 & 1.068 & 19.3\\
 \vspace{3pt}
      & PTF ($F_0=90$) & -8.023 & -7.748 & 0.275 & -10.080 & 1.166 & 1.166 & 20.3\\
  \hline
 \rule{0pt}{12pt}13.5
      & STF            & -8.577 & -8.377 & 0.201 & -10.275 & 0.949 & 0.949 & 16.1\\
      & PTF ($F_0=70$) & -8.754 & -8.554 & 0.200 & -10.286 & 0.866 & 0.866 & 15.5\\
 \vspace{3pt}
      & PTF ($F_0=90$) & -8.527 & -8.326 & 0.200 & -10.223 & 0.948 & 0.948 & 16.3\\
  \hline
 \rule{0pt}{12pt}13.9
      & STF            & -9.275 & -9.112 & 0.163 & -10.433 & 0.660 & 0.660 & 16.6\\
      & PTF ($F_0=70$) & -9.388 & -9.226 & 0.162 & -10.438 & 0.606 & 0.606 & 15.5\\
 \vspace{3pt}
      & PTF ($F_0=90$) & -9.229 & -9.066 & 0.164 & -10.386 & 0.660 & 0.660 & 16.4\\
\enddata
\end{deluxetable}

\begin{deluxetable}{ccccccccccc}
\tabletypesize{\scriptsize}
\tablecaption{Comparison between different phases for the
cases of $Y_p=0.3$ and $\rho_B = 10^{14.0}\,\rm{g\,cm^{-3}}$
at $T=1$ MeV and $10$ MeV.
The various quantities are as follows: $F$ is the free energy per baryon,
$S$ is the entropy per baryon, $\mu_n$ and $\mu_p$ are the chemical potentials
of neutrons and protons, $R_C$ is the radius of the Wigner--Seitz cell,
$n_n(r)$ and $n_p(r)$ are the number densities of neutrons and protons
at position $r$ in the cell.
\label{tab:3}}
\tablewidth{0cm}
\tablehead{
 \colhead{$T$} & \colhead{$\rm{phase}$} & \colhead{$F$} & \colhead{$S$} & \colhead{$\mu_n$} & \colhead{$\mu_p$} & \colhead{$R_c$} & \colhead{$n_n(0)$} & \colhead{$n_p(0)$} & \colhead{$n_n(R_C)$} & \colhead{$n_p(R_C)$} \\
 \colhead{($\rm{MeV}$)} & \colhead{} & \colhead{($\rm{MeV}$)} & \colhead{($k_B$)} & \colhead{($\rm{MeV}$)} & \colhead{($\rm{MeV}$)} & \colhead{($\rm{fm}$)} & \colhead{($\rm{fm^{-3}}$)} & \colhead{($\rm{fm^{-3}}$)} & \colhead{($\rm{fm^{-3}}$)} & \colhead{($\rm{fm^{-3}}$)}
 }
\startdata
 1 & Bubble  ($\rm{STF}$) & -9.615 & 0.152 & -0.678 & -40.265 & 17.13 & 0.0002 & 0.0000 & 0.0780 & 0.0392 \\
   & Droplet ($\rm{STF}$) & -9.528 & 0.155 & -0.026 & -31.934 & 18.28 & 0.0696 & 0.0263 & 0.0008 & 0.0000 \\
   & Droplet ($\rm{PTF}$) & -9.612 & 0.155 & 0.383 & -35.873 & 16.65 & 0.0807 & 0.0393 & 0.0013 & 0.0000 \\
 \vspace{3pt}
   & Uniform matter & -7.994 & 0.214 & -6.544 & -32.140 &  & 0.0422 & 0.0181 & 0.0422 & 0.0181 \\
  \hline
 \rule{0pt}{12pt}10
    & Bubble  ($\rm{STF}$) & -17.527 & 1.641 & -7.843 & -39.332 & 13.51 & 0.0124 & 0.0021 & 0.0559 & 0.0272 \\
    & Droplet ($\rm{STF}$) & -17.462 & 1.686 & -8.090 & -36.202 & 13.81 & 0.0547 & 0.0256 & 0.0275 & 0.0081 \\
    & Droplet ($\rm{PTF}$) & -17.481 & 1.661 & -7.881 & -37.735 & 14.56 & 0.0594 & 0.0297 & 0.0230 & 0.0061 \\
    & Uniform matter & -17.453 & 1.738 & -8.845 & -37.355 &  & 0.0422 & 0.0181 & 0.0422 & 0.0181 \\
\enddata
\end{deluxetable}

\clearpage

\begin{thebibliography}{}

\bibitem[Avancini et~al.(2009)]{Avancini09}
Avancini,~S.~S., Brito,~L., Marinelli,~J.~R., Menezes,~D.~P.,
Moraes,~M.~M.~W., Provid\^{e}ncia,~C., \& Santos,~A.~M. 2009, \prc, 79,
035804

\bibitem[Avancini et~al.(2010)]{Avancini10}
Avancini,~S.~S., Chiacchiera,~S., Menezes,~D.~P., \&
Provid\^{e}ncia,~C. 2010, \prc, 82, 055807

\bibitem[Blinnikov et~al.(2011)]{blin11}
Blinnikov,~S.~I., Panov,~I.~V., Rudzsky,~M.~A., \& Sumiyoshi,~K.
2011, \aap, 535, A37

\bibitem[Burrows et~al.(1984)] {burr84}
Burrows,~A., \& Lattimer,~J.~M. 1984, \apj, 285, 294

\bibitem[Burrows et~al.(2006)]{burr06}
Burrows,~A., Livne,~E., Dessart,~L., Ott,~C.~D., \& Murphy,~J. 2006,
\apj, 640, 878

\bibitem[Centelles et~al.(2007)]{TF07}
Centelles,~M., Schuck,~P., \& Vi\~{n}as,~X. 2007, Ann. Phys., 322, 363

\bibitem[De et~al.(2001)]{De01}
De,~J.~N., Vi\~{n}as,~X., Patra,~S.~K., \& Centelles,~M.
2001, \prc, 64, 057306

\bibitem[Furusawa et~al.(2011)]{furu11}
Furusawa,~S., Yamada,~S., Sumiyoshi,~K., \& Suzuki,~H. 2011, \apj,
738, 178

\bibitem[Furusawa et~al.(2013)]{furu13}
Furusawa,~S., Sumiyoshi,~K., Yamada,~S.,  \& Suzuki,~H. 2013, \apj,
772, 95

\bibitem[Gril et~al.(2012)]{Gril12}
Grill,~F., Provid\^{e}ncia,~C., \& Avancini,~S.~S.
2012, \prc, 85, 055808

\bibitem[Hempel \& Schaffner-Bielich(2010)]{hemp10}
Hempel,~M., \& Schaffner-Bielich, J. 2010, Nucl. Phys. A, 837, 210

\bibitem[Hirata et~al.(1996)]{hira96}
Hirata,~D., Sumiyoshi,~K., Carlson,~B.~V., Toki,~H., \& Tanihata,~I.
1996, Nucl. Phys. A, 609, 131

\bibitem[Janka et~al.(2007)]{jank07}
Janka,~H.-Th., Langanke,~K., Marek,~A., Mart{\'{\i}}nez-Pinedo,~G.,
\& M{\"u}ller,~B. 2007, Phys. Rep., 442, 38

\bibitem[Lattimer \& Swesty(1991)]{latt91}
Lattimer,~J.~M., \& Swesty,~F.~D. 1991, Nucl. Phys. A, 535, 331

\bibitem[Lattimer \& Prakash(2007)]{latt07}
Lattimer,~J.~M., \& Prakash,~M. 2007, Phys. Rep., 442, 109

\bibitem[Maruyama et~al.(2005)]{Maru05}
Maruyama,~T., Tatsumi,~T., Voskresensky,~D.~N., Tanigawa,~T., \& Chiba,~S.
2005, \prc, 72, 015802

\bibitem[Oyamatsu(1993)]{oyam93}
Oyamatsu,~K.
1993, Nucl. Phys. A, 561, 431

\bibitem[Oyamatsu \& Iida(2003)]{oyam03}
Oyamatsu,~K., \& Iida,~K. 2003, Prog. Theor. Phys., 109, 631

\bibitem[Schaffner \& Mishustin(1996)]{scha96}
Schaffner,~J., \& Mishustin,~I.~N. 1996, \prc, 53, 1416

\bibitem[Serot \& Walecka(1986)]{sero86}
Serot,~B.~D., \& Walecka,~J.~D.
1986, Adv. Nucl. Phys., 16, 1

\bibitem[G. Shen et~al.(2010)]{shen10}
Shen,~G., Horowitz,~C.~J., \& Teige,~S. 2010, \prc, 82, 015806

\bibitem[Shen et~al.(1998{\natexlab{a}})]{shen98a}
Shen,~H., Toki,~H., Oyamatsu,~K., \& Sumiyoshi,~K.
1998{\natexlab{a}}, Nucl. Phys. A, 637, 435

\bibitem[Shen et~al.(1998{\natexlab{b}})]{shen98b}
Shen,~H., Toki,~H., Oyamatsu,~K., \& Sumiyoshi,~K.
1998{\natexlab{b}}, Prog. Theor. Phys., 100, 1013

\bibitem[Shen et~al.(2011)]{shen11}
Shen,~H., Toki,~H., Oyamatsu,~K., \& Sumiyoshi,~K.
2011, Astrophys. J. Suppl., 197, 20

\bibitem[Sugahara \& Toki(1994)]{suga94}
Sugahara,~Y., \& Toki,~H. 1994, Nucl. Phys. A, 579, 557

\bibitem[Sumiyoshi et~al.(2005)]{sumi05}
Sumiyoshi,~K., Yamada,~S., Suzuki,~H., Shen,~H., Chiba,~S., \&
Toki,~H. 2005, \apj, 629, 922

\bibitem[Sumiyoshi et~al.(2008)]{sumi08}
Sumiyoshi,~K., \& R{\"o}pke,~G. 2008, \prc, 77, 055804

\bibitem[Sumiyoshi et~al.(2009)]{sumi09}
Sumiyoshi,~K., Ishizuka,~C., Ohnishi,~A., Yamada,~S., \& Suzuki,~H.
2009, \apj, 690, L43

\bibitem[Weber(2005)]{webe05}
Weber,~F. 2005, Prog. Part. Nucl. Phys., 54, 193

\end{thebibliography}
\end{document}